\documentclass[preprint]{aastex}
\newcommand{\etal}{{\it{et al.}}~}
\newcommand{\ie}{{\it{i.e.}}~}
\newcommand{\eg}{{\it{e.g.}}}
\newcommand{\Ha}{H$\alpha$~}
\newcommand{\ha}{H$\alpha$~}
\newcommand{\lha}{L$_{H\alpha}$}
\newcommand{\pccm}{pc cm$^{-6}$}
\newcommand{\arcs}{$^{\prime\prime}$}
\newcommand{\ang}{~\AA}
\slugcomment{To be published in ApJ, Oct 1, 2000.}

\begin{document} 

\title{The Contribution of Field OB Stars to the Ionization of the
Diffuse Ionized Gas in M33\footnote{Observations
made with the Burrell Schmidt Telescope of the Warner and Swasey
Observatory, Case Western Reserve
University.}~$^,$\footnote{Based on observations made
with the NASA/ESA Hubble Space Telescope, obtained from the data
archive at the Space Telescope Science Institute. STScI is operated by
the Association of Universities for Research in Astronomy, Inc. under
NASA contract NAS 5-26555.}}

\author{Charles G. Hoopes and Ren\'e A. M. Walterbos\altaffilmark{3}} 
\affil{New Mexico State University, Department of Astronomy, MSC4500, Box
30001\\ Las Cruces, New Mexico 88003}
\email{choopes@NMSU.edu,rwalterb@NMSU.edu}

\altaffiltext{3}{Visiting Astronomer, Kitt Peak National Observatory,
National Optical Astronomy Observatories, which is operated by the
Association of Universities for Research in Astronomy, Inc. (AURA)
under cooperative agreement with the National Science Foundation.}

\begin{abstract}

We present a study of the ionizing stars associated with the diffuse
ionized gas (DIG) and HII regions in the nearby spiral galaxy M33. We
compare our Schmidt H$\alpha$ image to the far-ultraviolet (FUV,
1520\ang) image from the Ultraviolet Imaging Telescope (UIT).  The
H$\alpha$/FUV ratio is higher in HII regions than in the DIG,
suggesting an older population of ionizing stars in the DIG. Assuming
ionization equilibrium, we convert the \ha flux to the number of
Lyman continuum photons N$_{Lyc}$. When compared to models of evolving
stellar populations, the N$_{Lyc}$/FUV ratio in HII regions is
consistent with a young burst, while the DIG ratio resembles an older
burst population, or a steady state population built up by constant
star formation, which is probably a more accurate description of the
stellar population in the field. The UIT data is complimented with
archival FUV and optical images of a small portion of the disk of M33
obtained with WFPC2 on HST. These images overlap low- and
mid-luminosity HII regions as well as DIG, so we can investigate the
stellar population in these environments. Using the HST FUV and
optical photometry, we assign spectral types to the stars observed in
DIG and HII regions. The photometry indicates that ionizing stars are
present in the DIG.  We compare the predicted ionizing flux with the
amount required to produce the observed \Ha emission, and find that
field OB stars in the HST images can account for 40\% $\pm$ 12\% of
the ionization of the DIG, while the stars in HII regions can provide
107\% $\pm$ 26\% of the \Ha luminosity of the HII regions. Due to the
limited coverage of the HST data, we cannot determine if stars outside
the HST fields ionize some of the DIG located in the HST fields, nor
can we determine if photons from stars inside the HST fields leak out
of the area covered by the HST fields. We do not find any correlation
between leakage of ionizing photons and \Ha luminosity for the HII
regions in our HST fields. However, the HST fields do not include very
luminous HII regions, and it would be worthwhile to see if there is
any trend at higher luminosities. If stellar photons alone are
responsible for ionizing the DIG, the current results are consistent
with no or few ionizing photons escaping from the galaxy.

\end{abstract} 


\section{Introduction}

An important component of the interstellar medium (ISM) in spiral
galaxies is the diffuse ionized gas (DIG, also called WIM for warm
ionized medium). DIG is a warm ($\sim$ 8000 K), diffuse
(n$_e$=0.1$-$0.2 cm$^{-3}$) layer of ionized hydrogen which permeates
the disks of spiral galaxies (see Walterbos \& Braun 1996 for a
review). In the Milky Way, where it is often referred to as the
Reynolds layer, DIG accounts for almost all of the ionized gas mass,
contributes about 30\% of the local HI column, and fills at least
20\% of the volume of the galaxy \citep{re91}. DIG in several edge-on
galaxies has been studied through \ha imaging
\citep{rkh90,det90,pbs94,r96,hwr99}, and the extent and brightness of
halo DIG has been found to vary dramatically. In face-on galaxies,
however, the most surprising discovery has been the {\it similarity}
of the DIG content among galaxies spanning a large range of
characteristics. The DIG components of many non-edge-on galaxies have
been studied: M31 \citep{wb94}, M33
\citep{hk90,g98}, M51 and M81
\citep{gwth98,whl97,whl99}, NGC 253 and NGC 300 \citep{hwg96}, NGC 247
and NGC 7793 \citep{f96}, and M101 \citep{whl97,whl99,g98}. Despite
the range of parameters seen in these galaxies, the contribution of
the DIG to the total \Ha luminosity (the {\it diffuse fraction}) is
consistently between 30 and 50\%. This is essentially a comparison of
the DIG luminosity and the HII region luminosity, so the constant
ratio establishes a link between the DIG and massive star formation.

Photoionization by massive stars and shock ionization by supernovae
have both been proposed as ionization sources of the DIG. All other
sources have been found to lack by far the required energy
\citep{r84}, although \cite{w98}
finds that cosmic ray ionizations may be a somewhat more significant
source than previously thought. The consistent diffuse fraction of 30
to 50\% of the total \Ha luminosity implies that a similar fraction of
the ionizing photons produced in a galaxy is required to power the
DIG, if photoionization is responsible for the DIG. The
diffuse fraction for M33 is 40\% \citep{g98}, which leads
to a minimum energy requirement (assuming all ionizing photons have
$\lambda$=912\ang) for the DIG of 1.8$\times$10$^{41}$ erg s$^{-1}$,
or 1.7$\times$10$^{-4}$ erg s$^{-1}$ cm$^{-2}$ of disk, using a radius
of 6 kpc for M33. This can be compared to 1.0$\times$10$^{42}$ erg
s$^{-1}$ and 1$\times$10$^{-4}$ erg s$^{-1}$ cm$^{-2}$ for the Milky
Way, using a radius of 15 kpc, which just equals the amount of energy
provided by supernovae \citep{r84}. In M33 the
supernova rate is 1/360 years \citep{gklbds98},
which leads to 0.9$\times$10$^{41}$ erg s$^{-1}$, using the canonical
10$^{51}$ erg for each supernova. This is a factor of 2 too low to
account for the DIG, even assuming an unrealistic 100\% efficiency in
converting the energy into ionized gas. Also, the
\Ha luminosity used here has not been corrected for internal
extinction, which would make the power requirement even
higher. Although some DIG may be shock ionized, supernovae cannot be
the source of energy responsible for the bulk of the ionization of the
DIG in M33 or the Milky Way, through shock ionization or any other
source which taps supernovae energy, such as turbulent mixing layers
\citep{ssb93}.

This leaves OB stars as the most likely source. Photoionization is
consistent for the most part with spectroscopic observations of the
DIG. Emission from [SII] 6716, 6731\AA~ and [NII] 6548, 6584\AA~ is
enhanced relative to \ha in the DIG when compared to HII regions, and
[OIII] 4959, 5007\AA~ is fainter relative to H$\beta$
\citep{r97,gwb97,hrt99,r85}. This spectrum has been reproduced by
photoionization models \citep{dm94} using a dilute radiation field in
a diffuse medium. Low limits on HeI 5876\AA/\ha in the Milky Way
\citep{rt95} have yet to be explained, and in NGC 891 the HeI/\ha
ratio, while higher than in the Milky Way, is still too low to be
consistent with other line ratios \citep{r97}. In M31, however, the
HeI emission is consistent with the Domgorgen \& Mathis models
\citep{gwb97}, at least for the brightest DIG, and the HeI/\ha ratio
in several irregular galaxies is also consistent with photoionization
\citep{mk97}. A more challenging problem has been posed by high
[OIII]/H$\beta$ ratios seen in the DIG of several
galaxies \citep{whl97,r98}, which still have not been reproduced by
pure photoionization models. Nevertheless, the bulk of the evidence
suggests that OB stars are the dominant ionization source for the DIG.

There is still uncertainty, however, regarding the location of the OB
stars responsible for ionizing the DIG. Are the ionizing photons
leaking out of density-bounded HII regions, or are there enough {\it
field OB stars} to provide the ionizing photons?  Either option is a
deviation from commonly held views of the ISM. If field stars are the
dominant source of ionizing photons, this requires a substantial
population of massive stars outside of HII regions.  Do these stars
form in the field, do they manage to drift out of the cloud in which
they formed, or is the gas swept away by the supernovae of even more
massive stars? On the other hand, if the photons which are ionizing
the DIG are leaking out of HII regions, an appreciable population of
density-bounded HII regions is required. HII regions are usually
treated as being radiation-bounded, which makes the determination of
star formation rates from \Ha luminosities straightforward. Does such
a population of leaky HII regions exist?

Evidence for field OB stars does exist
\citep{tlp74,hs80,gcc82,m95a}. \cite{m95a} defined field
stars as those further from an HII region than a star could travel in
the lifetime of an OB star ($<$ 10$^7$ years), a few tens of parsecs
at most. Using ground-based UBV photometry, \cite{pw95} investigated
the distribution of OB stars in M33. They concluded that 50\% of the
ionizing stars were found outside of HII regions. Such a study is
difficult using ground-based optical data, where both crowding and the
color degeneracy of hot stars pose serious problems. A systematic
search for field OB stars is needed, and space based photometry in the
ultraviolet can provide a strong test of the optical results.

The alternative to field OB stars is a significant population of
density-bounded HII regions. Individual HII regions which are
density-bounded do exist, such as the Orion nebula \citep{rshe91} and
the starburst in NGC 4214 \citep{lwcfrs96}. \cite{ok97} compared the
\Ha luminosities of several HII regions in the LMC with the predicted
ionizing fluxes of the stars within. They found 2 out of 14 HII
regions to be substantially leaking ionizing photons, and 3 others
that were slightly leaky. The observed absence of Lyman continuum
radiation leaking from spirals \citep{lfhl95} suggests that such
radiation is contained within HII regions, although it is conceivable
that HII regions are just leaky enough to ionize the DIG, but not so
much that an appreciable number of ionizing photons escapes the entire
galaxy \citep{whl97}. There is evidence from the \ha luminosity
functions of HII regions in spirals that high luminosity HII regions
may be density-bounded \citep{b00}.

Here we focus on identifying the stars responsible for ionizing the
DIG. In order to isolate massive stars we turn to the far-ultraviolet
(FUV), where OB stars dominate the stellar emission.  The local
group spiral M33 is an ideal galaxy to study the ionization of the
DIG, due to its low inclination, vigorous star formation rate, and
distance of only 0.84 Mpc \citep{f91}. In section 2 we discuss the
data used to carry out this project, which include
\Ha imaging of M33, FUV imaging using the Ultraviolet Imaging
Telescope (UIT), and optical and FUV stellar photometry from the Wide
Field Planetary Camera 2 (WFPC2) on {\it Hubble Space Telescope}
(HST). In section 3 we briefly review the properties of the DIG in
M33. In section 4 we discuss the FUV emission on large scales from the
UIT image and its relation to the \Ha emission from both DIG and HII
regions. In section 5 we present the results of the stellar photometry
in HII regions and DIG. In section 6 we compare the FUV information
from HST and UIT. Section 7 contains a discussion of the results and
conclusions.

\section{The Data}

\subsection{The \ha Mosaic}

Table 1 contains a summary of the observations. The ground-based data
consist of H$\alpha$ and off-band continuum images of M33 obtained
with the 0.6 meter Burrell-Schmidt telescope at Kitt Peak National
Observatory in November 1995 \citep{g98}. The 2048$^2$ detector
provided a field of view of 1.15$\times$1.15 degrees$^2$ at
2.028\arcs/pixel. The images were offset using the shift-and-stare
technique to minimize flat-field irregularities. The final mosaic
covers an area of about 1.75$\times$1.75 degrees$^2$. We reduced the
images using standard procedures in IRAF. Twilight flats from all
nights of the observing run were co-added to create a ``super flat''
which was used to correct for gain variations. We calibrated the
continuum image using the published R magnitude and the known shape
and transmission of the continuum filter. The H$\alpha$ image was
calibrated relative to the continuum, again using the shape and
transmission of the line filters. The continuum image was scaled to
the H$\alpha$ image using foreground stars and subtracted. The total
H$\alpha$ flux we observed is 3.6 $\times$ 10$^{-10}$ ergs s$^{-1}$
cm$^{-2}$. Spectroscopy of DIG and HII regions in M33 (Hoopes \&
Walterbos 2000, in preparation) shows that [NII] 6584\AA~ emission is
about 20\% as strong as \ha in both the DIG and HII regions. The lack
of [NII] enhancement in the DIG and the lower ratios in both
environments than seen in other galaxies are probably due to a lower
Nitrogen abundance in M33 \citep{v88}. The result is that at most only
5\% of the light detected in the \ha filter can arise from [NII]
emission, and the amount of contamination is the same for both HII
regions and DIG. The continuum-subtracted \Ha image is shown in figure
1.

\subsection{The UIT Image}

We obtained an archival UIT image of M33 taken on the {\it Astro-1}
space shuttle mission \citep{s92,s97}. We used the FUV image, a 424
second exposure taken through the B1 filter ($\lambda$=1520\AA,
$\Delta\lambda$=354\AA). The image was originally produced on film and
later digitized. The resolution of the image is about 3\arcs and the
field of view is roughly 40$^{\prime}$. The \Ha image was convolved to
similar resolution and put on the same grid. UIT was calibrated in the
laboratory before launch and also in flight by comparing UIT images
with previous UV space missions. The calibration uncertainty is about
15\% \citep{s92}. The UIT image can be found in \cite{l92}.

\subsection{WFPC2 Images and Stellar Photometry}

Lastly, we used archival HST WFPC2 images of regions in M33, from
project GO6038. These images were originally obtained to study hot
stars and star clusters in M33 \citep{cbf99}. Figure 1 shows that the
pointings are restricted to the inner $\sim10^{\prime}$ radius of M33,
but all of the pointings contain HII regions and DIG. The pointings
near the center cover regions that are more crowded than those covered
in the pointings at larger radii. Spiral arms and inter-arm regions
are both sampled. The only obvious bias introduced by the locations of
the images is that they do not include the brightest HII regions in
M33. The filters and exposure times are F170W (2$\times$900 s), F336W
(2$\times$900 s), F439W(600 s) and F555W(160 s). The two separate
exposures in the UV and FUV filters were combined to remove cosmic
rays. The images were corrected for the $\sim$ 4\% charge transfer
efficiency problem of the WFPC2 by multiplying with a ramp image as
described by \cite{h95}. The F170W and F336W images were also
corrected for degradation of throughput after decontamination, again
using the formula in \cite{h95}.

Photometry was performed on the HST images using the DAOPHOT
package. For each frame, a PSF model was created using bright,
isolated stars. The empirical PSF model was found using an iterative
process. A constant PSF model was first computed and used to subtract
the neighbors of the PSF stars. Then a linearly varying PSF model was
computed on the image with the PSF neighbors subtracted, and then used
to remove the PSF neighbors from the original image. Finally a
quadratically varying model was computed from the PSF stars with the
neighbors subtracted. If there were few PSF stars (less than about 15,
which was usually the case in the F170W images) a constant psf model
was used. It was also created iteratively, first making a model PSF,
then subtracting any neighbors of the PSF stars and revising the PSF
based on the subtracted image. We compiled a list of stars detected in
all four bands. The magnitudes were put on the STMAG photometric
system, described by \cite{h95}, using the most recently determined
calibration for WFPC2. The calibration uncertainty is 1$-$2\% in the
optical filters, and around 10\% in the F170 filter.

\section{DIG in M33}

The \Ha mosaic of M33 is discussed in detail in \cite{g98}. We briefly
repeat the relevant results here. The diffuse fraction of M33 was
found to be 40\%, in good agreement with all other non-edge-on spirals
studied so far. The Schmidt image reveals the DIG to be a complex
network of filaments and arcs, superimposed upon a fainter, more
diffuse background. As in other spirals, DIG is concentrated in
regions of high star formation, and in the inner disk, but it is much
more extended than the star forming regions, filling the disk with
faint \Ha emission. Narrow-band imaging of the DIG in M33 \citep{g98}
reveals that it shows the same spectral signature of enhanced [SII]
seen in other galaxies.

\section{H$\alpha -$UIT Analysis}

The FUV/\ha ratio in the DIG can constrain the source of
ionization. For example, if the measured ratio in the DIG did not
correspond to any reasonable average spectral type or population of
ionizing stars, we would conclude that the ionizing photons in the DIG
are not produced locally. To perform this test we measured the FUV
flux (L$_{1520}$) in the UIT image and the \ha flux in the Schmidt
image. For HII regions, we defined an aperture and a background
annulus on the \ha image, and then used the same aperture and annulus
on the FUV image. For DIG, we measured the fluxes in 500 pc $\times$
500 pc square apertures (in the plane of the sky), with the HII
regions masked out. The HII region mask was defined on the \ha image
and applied to both the \ha and FUV images. This is slightly different
from the approach taken in \cite{hw97}. In that paper the fluxes were
measured in smaller apertures, leading to the possibility that in HII
regions an aperture may miss the concentrated cluster of OB stars
while still containing \ha emission, leading to very high
\ha/L$_{1520}$ ratios and thereby increasing the scatter in the
distribution of ratios. Also in that paper no background was
subtracted from the HII regions, contrary to the approach taken
here. The background subtraction typically affects the \ha flux of the
HII regions by only 1$-$2\%, but in the FUV image the contrast between
HII regions and the field is not as high as in H$\alpha$, and the background
subtraction can reduce the FUV flux by as much as 50\%, and even more
for faint HII regions.

We computed the number of ionizing photons (N$_{Lyc}$) from the \Ha
luminosity, assuming ionization equilibrium and case B recombination
\citep{o89} and neglecting absorption of ionizing
photons by dust. The histograms in figure 2 show the resulting
distribution of N$_{Lyc}$/L$_{1520}$ ratios for the two
environments. The histograms are shown sideways for comparison with
models (see below). The observed ratios are shown; they have not been
corrected for foreground or internal extinction. The
N$_{Lyc}$/L$_{1520}$ ratio in DIG is lower than that in HII
regions. This fits a scenario in which the ionizing stars in the DIG
are of later type than those in HII regions, since the later type
stars would produce a lower ratio of ionizing to non-ionizing UV
photons.

In figure 2 we compare the N$_{Lyc}$/L$_{1520}$ ratios with models of
evolving stellar populations. We used the Starburst99 evolutionary
model to compute the ratio of ionizing photons to FUV flux
\citep{l99}. The model used a \cite{s55} initial mass function (IMF)
with an upper mass cutoff of 120 $M_{\odot}$. LMC-like (Z=0.008)
metallicity models were calculated. The metallicity of M33 in the
central region is about solar, but there is a steep gradient towards
lower metallicity with distance from the center \citep{v88}, so that
overall M33 is subsolar in metallicity. Single burst and steady state
models are presented. The reddening bars shown indicate the
direction that the {\it models} would move if reddened by the given
amount. Both the LMC \citep{h83} and Galactic \citep{ccm89} extinction
laws are given, and the case of a foreground screen and a uniform
mixture of dust and gas are shown for each extinction law. \cite{m95b}
found an average color excess of E(B-V)=0.16 in several fields of
M33. The Galactic foreground produces a color excess of
E(B-V)=0.03$\pm$0.02 at this location \citep{mr69}, and the rest is
internal to M33.  The extinction in the \cite{m95b}) fields ranges
from E(B-V)=0.09 to 0.33. We also show reddened versions of each
model. The steady state model, which represents the population
expected outside of HII regions, was reddened by E(B-V)=0.1, while the
single burst model, which represents the population in an HII region,
was reddened by E(B-V)=0.2, all using the LMC extinction law. The
dashed line is reddened assuming a uniform mixture, and the dotted
line is reddened assuming a foreground screen. Each reddened model
also includes E(B-V)=0.03 of foreground Galactic reddening, using the
Galactic extinction law.

The DIG distribution is consistent with an older burst population, or
with a steady state, constant star formation model. The steady state
model is probably a more accurate description of the stellar
population in the DIG, but it is only an approximation. Figure 2 also
shows the predictions of the steady state model using steeper IMF
slopes than the usual Salpeter slope of $\alpha$=$-$2.35. Although the
predicted ratios are below the observed ratio, with a reasonable
amount of extinction the prediction for $\alpha$=$-$3.0 would match
the observed DIG values well. This is interesting in light of the fact
that \cite{m95a} found a steeper IMF slope for OB stars in the field
in the LMC and SMC. We will explore this issue further in section 6.
The HII regions resemble a young burst population. The model values
are dependent on the assumed IMF, and in small HII regions (which are
the most numerous) that have low numbers of massive stars, the
ionizing flux of a single O star may dominate the light from the UV
emitting B stars.  Figure 3 shows the same histograms along with the
expected ratios from ionizing stars, taken from the CoStar stellar
models \citep{sd97}. We show models using low metallicity (Z=0.004)
and solar metallicity (Z=0.020), to cover the extreme values. Again
the DIG ratio can be explained by later type ionizing stars.

The N$_{Lyc}$/L$_{1520}$ ratio in the DIG can be explained with a
local ionization model, where the ionizing population in the DIG is
older than that in HII regions. However, the UIT analysis is far from
conclusive. The ionizing photons which produce the \ha emission seen
in the DIG may be produced elsewhere and then leak into the DIG, as in
the leaky HII regions explanation for the ionization of the DIG. In
this case the models in figure 2 and 3 would not apply. To test this
more conclusively, we need to find out whether the ionizing photons
are produced locally. We investigate this further in the next section.

If the difference in the FUV/\ha ratio between DIG and HII regions
were due to higher extinction in the HII regions, at least 0.9
magnitudes A$_V$ in excess of that in the DIG would be necessary to
explain the result using a foreground screen model, and much higher
for a uniform mixture. \cite{gwb97} found, on average, a 0.3
magnitude difference in extinction between HII regions and DIG in
M31. Our stellar photometry (see the next section) indicates that
there is on average little difference in the amount of extinction in
the two environments.  However, this is for only a very small part of
the disk, so to be consistent with \cite{gwb97} we reddened the
burst models more than the steady state models in figures 2 and 3.

\section{HST Photometry}

\subsection{Spectral Classification of OB Stars}

The most conclusive test of whether field stars can ionize the DIG
would be to find the spectral types of all the OB stars outside of HII
regions, add up the number of ionizing photons contributed by each
star, and compare the total to that required by the \Ha luminosity of
the DIG. \cite{m85} and \cite{m95b} have shown that optical photometry
is unable to distinguish between the different spectral types of
ionizing stars, a distinction that must be made if accurate Lyman
continuum luminosities are to be determined. This is because the peak
of the Planck spectrum lies blueward of these filters for stars of
30,000 K or hotter, so optical filters sample the Rayleigh-Jeans
portion of the spectrum which has a constant slope independent of
temperature. Spectroscopy is the most reliable way to determine
spectral types of massive stars, but obtaining spectroscopy of a large
sample of field stars in M33 would require a prohibitive amount of
telescope time. However, FUV photometry combined with optical
photometry can be used to estimate spectral types of ionizing stars
with reasonable accuracy. The F170W is closer to the peak of the
Planck spectrum for temperatures above 30,000 K, providing an
advantage over optical photometry alone.

The archival HST images cover DIG of varying brightness (see overlay
of figure 1). However, they do not overlap any very luminous HII
regions, only moderate and low-luminosity HII regions. This is a
limitation of the data, and it prevents us from investigating whether
the brightest HII regions are density-bounded. It should be noted that
the HST images cover a much smaller area of the disk of M33 than the
UIT image, so the focus of the analysis shifts here from a global to a
local perspective.  The HST pointings are all located in the inner
disk of M33, within 10$^\prime$ from the nucleus, where the
metallicity is close to solar \citep{v88}. For this reason we assume
solar metallicity when employing stellar models in the analysis of the
WFPC2 images.

We put the \Ha image of M33 on the same grid as the WFPC2 frames, and
then classified the stars in the WFPC2 images as being either in an
HII region or in the field based on the \Ha surface brightness. An
isophotal cut was used, but the level was adjusted to satisfactorily
isolate the two environments based on their morphology. In the inner
disk the cut level is as high as 200 \pccm, while in the outer disk it
is as low as 60 \pccm. Individual HII regions were then picked from
the masked image by eye. The rest of the image was counted as
DIG. Only the Wide Field images were used, as there were no complete
HII regions in the Planetary Camera images, and they do not cover a
large enough area of DIG to analyze.

\cite{h95} caution that the response curves
of the main photometric filters used with WFPC2 (F336W, F439W, F555W),
are sufficiently different from the groundbased analogs (UBV), that
any reddening correction must be applied to the photometry in the HST
filter system. Thus we derived reddening relations for the HST filter
system used in this project, rather than transform the photometry to
the groundbased system. We used the model stellar spectra of
\cite{lcb97}, and reddened them according to both the Galactic
extinction law \citep{ccm89} and the LMC law \citep{h83}. We used the
\cite{lcb97} model spectra rather than the CoStar models used below
because the Lejeune \etal models extend to spectral types later than
B0, allowing more of the unevolved main sequence to be used in
determining the extinction. The reddened spectra were multiplied by the
HST filter response curves to obtain the magnitude. The derived
relations agree with those found in \cite{h95}, although the F170W
relation was not given. We found that the LMC extinction law worked
more satisfactorily than the Galactic law for reddening all bands to
match the observed colors. As before, the reddening includes a
correction for E(B-V)=0.03 foreground reddening using the Galactic
extinction law, and any excess reddening using the LMC extinction law.

Figure 4 shows representative color-magnitude diagrams (CMDs) for HII
regions and DIG regions in M33. Isochrones of ages 0 and 7$\times10^6$
years are also shown. The isochrones were generated using the Geneva
stellar evolution models \citep{s93}. In order to put the isochrones
in the STMAG system, for each stellar mass at each age we found the
closest match in T$_{eff}$ and $log g$ in the model stellar spectra of
\cite{lcb97}. The spectrum was then multiplied by the WFPC2 filter
response curves to derive the observed magnitude. The isochrones are
shown for reference only, they are not the best fit to the CMD. The
purpose of these isochrones was to determine the extinction for each
region. The amount of extinction needed to match the observed main
sequence with the isochrone is taken as the average extinction for
that region. The same was done for the remaining stars not in HII
regions, which are field stars in the DIG. Using a single value for
extinction in either environment is obviously an over-simplification,
but it is the best solution for the available data.  The average
extinction found in HII regions was E(B-V)=0.14, and in DIG it was
E(B-V)=0.13. We estimate that the extinction can be determined with
this method with a 1-$\sigma$ accuracy of E(B-V)= $\pm$0.02.

We assigned spectral types based on the CoStar models of
\cite{sd97}. They provide spectral energy distributions for stars of
mass 20, 25, 40, 60, 85, and 120 M$_{\odot}$ at various ages in the
\cite{l96} database. We used the solar metallicity (Z=0.020) models to
match the metallicity where the HST pointings are located.  We ran the
model spectra through the HST filters, and linearly interpolated the
magnitudes and other stellar properties (\eg~ temperature, surface
gravity, and ionizing photon luminosity) to the ages and masses
between those given, to create a complete grid with 1 M$_{\odot}$ mass
resolution and 10$^5$ year time resolution. We then searched for the
nearest match to the four magnitudes measured for each star. Thus for
each star we go directly from the measured photometry to ionizing
luminosity. The uncertainty in the ionizing luminosity is determined
by the uncertainty in the photometry, ignoring uncertainty in the
accuracy of the models, which is difficult to quantify.

We then inspected the color magnitude diagrams and the positions of
the stars that were classified, noting any UV-bright stars that were
not classified. In order to classify these stars, we adjusted the
tolerance for the F555W magnitude, and if necessary for the F439W
magnitude as well. In most cases the tolerance was increased to
2-$\sigma$, and in a few cases, 3-$\sigma$. A small number of stars
were in crowded regions and obviously had mismatched F555W and F439W
magnitudes (crowding usually does not affect the UV magnitudes since
fewer stars show up in the UV images), so we used only F170W and F336W
magnitudes for classification. Most stars were matched within
1-$\sigma$ of the four measured magnitudes.

We compared our photometric spectral types to previous spectroscopic
spectral types to assess the accuracy of the photometric technique. We
looked for overlapping stars in \cite{m95b} and
\cite{m96}, finding only two stars that
overlap, both in \cite{m96}. The reason is that our technique only
assigns spectral types to B0 stars and earlier, so while many B stars
overlapped, we did not assign spectral types to them, so they could
not be used for comparison. The properties that the photometric
technique assigned to the two stars are shown in table 2, along with
the properties of the spectral type assigned by \cite{m96}. These
properties are taken from \cite{vgs96}. The properties that the
photometric technique assigned to both stars match reasonably well
with the properties expected for stars of the correct spectral
type. It should be noted that UIT104 is a Wolf-Rayet star, and might
not be expected to match the properties of an O9Ia star. Nevertheless,
the agreement gives us reason to believe that the photometrically
determined spectral types agree with those determined
spectroscopically, to within about a spectral type.

\subsection{Comparison of Predicted and Observed \ha Luminosity}

For each HII region we summed the predicted ionizing luminosity of all
the stars included in the region. The predicted ionizing luminosity in
the DIG was the cumulative luminosity of stars not included in an HII
region. Each chip of each WFPC2 pointing was considered separately. We
converted the predicted ionizing luminosity to a predicted \ha
luminosity \lha, assuming that all ionizing photons result in an
ionization (\ie no dust absorption) and case B recombination. The
predicted \lha~ was then compared to the observed \lha. We measured the
observed \lha~ from the \ha image (figure 1). The boundaries of the HII
regions were determined on the masked image, and a background
correction for the surrounding DIG was subtracted. HII regions which
were not completely contained within the HST pointing were not
considered. The \lha~ for DIG was the sum of all the luminosity not in
an HII region. The luminosities were corrected for extinction using the
estimates of E(B-V) from the color-magnitude diagrams.

In figure 5 we compare the predicted \lha, based on the ionizing
luminosity of the stars present, with the observed \lha~ from the
\Ha image for both HII regions and DIG. The errors reflect a combination
of uncertainty in photometry and in extinction. The effect of
photometric errors on the ionizing luminosity was calculated by
determining for each star the model with the highest N$_{Lyc}$ and the
model with the lowest N$_{Lyc}$ that still matched within the
photometric errors. When the best match luminosities were added for
all stars in the region, the range found for each star was added in
quadrature. The effects of uncertainty in extinction were quantified
by varying the correction by E(B-V)=$\pm$0.02, and again calculating
the total ionizing luminosity of the stars in a region. These two
uncertainties were added together in quadrature to produce the
error bars shown in figure 5.

Qualitatively, we find that the measured \lha~ for both HII regions
and DIG agree well with the predicted luminosities. There is
considerable scatter, but given the uncertainties in stellar models,
extinction in the UV, and spectral classification without the use of
spectra, the agreement is encouraging. The observed \lha~ is corrected
for extinction, but the predicted \lha~ may be an overestimate since
we ignore the possibility of ionizing photons being absorbed by
dust. To correct for this we would have to know how much of the
ionizing luminosity of a star is absorbed by dust. \cite{mw97}
estimate the fraction of dust absorbed in Galactic HII regions to be
about 25\%, perhaps slightly lower for smaller HII regions such as
those we consider here.

Figure 5 shows that while the mean ratio is close to unity, some HII
regions are under-predicted or over predicted. This is an indication
that predictions for individual regions are not reliable, but that the
average for all the regions is better. Thus a comparison of the
average values for DIG and HII regions is more illuminating than
individual regions.  The average ratio of predicted to required \lha~
are given in table 3. For HII regions the predictions are consistent
with the observed \lha. For DIG regions there are not enough ionizing
photons emitted by field stars to account for all of the observed \ha
emission, but there are enough to provide a significant fraction of
the ionization, about 40\%.

To test the idea that FUV information is necessary to estimate the
spectral types of ionizing stars, we re-classified the ionizing stars
using only the F336W, F439W, and F555W magnitudes. The results are
given in table 3. The predicted \lha~ is higher, and most HII regions
are overpredicted, leading to a large excess of ionizing photons. If
the FUV information had been ignored in the initial classification,
many non-ionizing stars would have been classified as ionizing stars,
leading to a further excess. The FUV information is crucial to
accurately predict ionizing fluxes, and optical information alone will
lead to a severe over prediction of ionizing fluxes of OB stars. This
test underscores the difficulty of drawing conclusions about OB stars
based on optical photometry (Patel \& Wilson 1995, see also O'Dell,
Hodge, \& Kennicutt 1999). This may account for the discrepancy in the
fraction of OB stars found outside of HII regions. We identified 116
ionizing stars in the five WFPC fields. Of these, 27\% (31 stars) were
in the DIG, lower than the $\sim$50\% found by \cite{pw95}. However,
it is important to remember that we can only investigate a small
fraction of the disk of M33 with the present HST data.

\subsection{Characteristics of the HII Regions}

\cite{ok97} found that 2 out of 14 HII
regions in the LMC are leaking a sizeable number of ionizing
photons. These two HII regions both have a ring-like morphology,
leading to the question of whether HII morphology is correlated the
amount of leakage. This is an important question in light of
suggestions that ionizing photons escaping from superbubbles may be an
important ionization source \citep{dsf00}.  Softening of the spectrum
by re-radiation of ionizing photons through chimney walls may explain
the low HeI measurements \citep{n91}. We classified the HII regions in
the HST field based on their morphology in the \Ha image, in order to
look for trends in the amount of leakage with morphology. Figure 6
shows this comparison. The HII regions are either compact, meaning
center brightened and obviously discrete, or diffuse, which includes
rings, filaments, and faint tenuous HII regions (the ring-like HII
regions are shown as double open circles). It appears that compact and
diffuse HII regions are equally likely to be density-bounded.

In figure 6 we also see that there is no significant trend for
leakiness with \lha. It is very important to remember that the
brightest HII regions in M33 were not imaged in the HST data, and only
a few in our sample are more luminous than 10$^{38}$ erg s$^{-1}$
(corrected for internal extinction).  Thus we cannot address the issue
of whether the most luminous HII regions are density-bounded
\citep{b00}. It is still interesting, however, that the two brightest
HII regions in M33, NGC 604 \citep{gp00} and NGC 595 \citep{mwp96},
appear to be radiation-bounded. In a future paper we will investigate
the brightest HII regions in M33.

\section{UIT - HST Comparison}

The WFPC2 images resolve stars above a limiting magnitude, which in
the F170W filter corresponds to about a B0 star. The UIT image
measures all of the FUV surface brightness of the entire population of
UV emitting stars. The UIT image sensitivity extends to 50-100 Myr old
populations \citep{o97}, which corresponds to B and early A type stars,
so a comparison of the UV flux measured by HST and UIT thus compares
the stars of spectral type B0 and earlier to the entire OBA
population. This is an age dependent ratio which can be modeled, and
may be used to further characterize the stellar population in both HII
regions and DIG. The attractive feature of this comparison is that it
is largely independent of extinction, as both filters cover a similar
range of wavelengths (however extinction must still be incorporated,
see below).

We have performed this comparison for the HII regions and DIG on the
HST fields. The FUV flux from the UIT image was measured using the
same apertures used to measure the \Ha flux. We then summed up the FUV
flux of the stars brighter than a magnitude limit in the same regions
in the HST images. In order to keep a consistent limiting magnitude,
we reddened the F170W magnitude of a B0 star (M$_{F170W}$=16.92) using
the extinction found from the CMDs and used throughout this analysis,
and summed the flux from all stars brighter than this magnitude. This
is the point where extinction becomes important in this comparison,
the determination of the cutoff magnitude in the HST images. The
relationship between flux and magnitude in the STMAG system is given
by
\begin{equation}
M_{F170W}=-2.5 log (F_{\lambda}) -21.1
\end{equation}
\citep{h95}. The flux was determined from the
magnitudes {\it without} correcting for extinction, as the UIT flux is
also uncorrected. The extinction is only used to determine the cutoff
magnitude which corresponds to a B0 star.

A histogram of the ratios is shown in figure 7.  The histogram shows a
difference between the two populations, with DIG having a lower ratio
of HST to UIT flux. This is expected if the DIG is ionized by an older
population, with fewer stars earlier than B0 relative the number of
stars later than B0. Also shown on the plot are models of the
evolution of this ratio through time, constructed using Starburst99
\citep{l99}. The UIT flux was modeled as an evolving population with a
mass range from 1 to 120 M$_\odot$, a \cite{s55} IMF, and solar
metallicity.  In order to simulate the populations seen in the HST FUV
images, we needed to include flux only from stars with
M$\ge$20M$_\odot$, as this corresponds to the limiting spectral type
of B0 in the HST images. To do this we ran the models again, except
that mass range was 20 to 120 M$_\odot$, and the total mass was scaled
down by 0.2 (so that the total mass, including stars with
M$\le$20M$_\odot$, was 10$^6$M$_\odot$, as it was for the UIT
simulation). This approach is simplistic, but provides a first-order
model for comparison with the observations. As expected, the ratio of
HST to UIT flux in the single burst model generally decreases as time
passes and the most massive stars die. The observed ratios for HII
regions match well with the predicted values for single burst
populations. An older burst model reproduces the DIG ratios reasonably
well.

We also ran steady state models as described above, but the resulting
ratio did not agree well with the observed ratio for the DIG, contrary
to the analysis in figure 2. The disagreement may indicate that if the
steady state model is an accurate description of the stellar
population in the DIG, the parameters used for the HII region stars
may not apply to the field stars. Specifically, \cite{m95a} derived
the IMF for field stars in the LMC and SMC, and found a slope ranging
from $\alpha=-4.7$ to $-5.1$ (in the notation where the Salpeter IMF
slope $\alpha$ is $-2.35$). We ran models using IMF slopes for massive
stars (M$\ge$20M$_\odot$) of $\alpha$=$-3.5, -3$, and the standard
Salpeter slope of $-$2.35, while keeping the slope for less massive
stars at $\alpha$=$-2.35$. The observed HST/UIT flux ratio agrees much
better with the models using a steeper slope of $\alpha$=$-3$ than the
Salpeter slope. This supports the conclusions of \cite{m95a} regarding
the IMF slope for field OB stars. The observed ratios suggest that the
IMF slope for field OB stars in M33 is not as steep as that in the LMC
and SMC, but is still steeper than the IMF in HII regions.

It should be noted that the Starburst99 models are intended for
starburst regions, \ie regions with a large number of stars. Most of
the HII regions we are investigating do not fit into this category, so
the model cannot be expected to accurately predict the observed
ratio. The main problem is that with small numbers of massive stars,
the IMF is not well sampled. This may be the reason behind the spread
in the observed distribution of HST/UIT ratios, and also might explain
why some HII regions have ratios too high to be explained by the
models. However, the predicted ratio agrees with the majority of HII
regions, which suggests that these deviations average out for a large
number of HII regions. Also note that the two filters are not
identical, and this is probably the reason some of the HII regions
have HST/UIT $>$1.

\section{Discussion}

Our most important result is that the OB stars outside of HII regions
can account for 40\% of the ionization of the DIG in M33. There are
several points to keep in mind. One is the scale we are investigating,
which is constrained by the size of the Wide Field chips of WFPC2, 325
pc across each chip. This means we are implicitly assuming that stars
(field stars in particular) do not have an influence at distances
greater than this, or that the number of photons escaping the image is
balanced by the number of photons coming in from outside the image. In
a uniform medium with n=0.2 cm$^{-3}$, which is the density found for
the DIG in the Galaxy, the Str\"omgren sphere of an O3V star has a
diameter of about 770 pc (Osterbrock 1989, using the stellar ionizing
flux in Vacca \etal 1996). However there are few of these stars
outside of HII regions. One chip can completely contain the
Str\"omgren sphere of an O8V star, and a B0V star ionizes a region
about 170pc across. The density in the DIG in the inner regions of M33
may be higher than the density measured in the solar neighborhood,
which would tend to make the ionized regions smaller. We cannot
account for neighboring HII regions or ionizing stars outside of the
WFPC2 field of view.

Another point to consider is the many uncertainties dealt with in this
analysis. When working in the UV extinction is always a prime concern,
and small changes in the adopted extinction, or in the extinction law,
can cause large changes in the ionizing luminosity. When dealing with
individual stars in external galaxies, crowding can often be a
concern. The stars visible in the F170W filter are rarely crowded, but
of course crowding may be present but not detectable, as in the case
of binaries. Crowding may also be a more severe problem for OB
associations in HII regions than for the more sparsely distributed
stars in the DIG. Another source of uncertainty stems from the stellar
atmosphere models which we use to assign Lyman continuum
luminosities. The stellar models give excellent agreement for the
non-ionizing spectrum of massive stars, but the ionizing spectrum is
difficult to test without making assumptions. We have tried to be as
conservative as possible in every aspect of this analysis. Given these
uncertainties, it is remarkable that we find such good agreement
between the predicted and observed \lha. This gives us confidence in
the stellar models, and it also suggests that any mistakes we are
making regarding the extinction and spectral classification are
relatively small. It also suggests that we are detecting most, if not
all, of the ionizing stars.

Without placing trust in the absolute accuracy of the predicted
ionizing fluxes, it is possible to draw conclusions by comparing our
predictions for DIG and HII regions. This comparison relies on the
assumption that any errors in the prediction for the DIG will also be
made for HII regions. The ratio of N$_{Lyc}$/\lha~ is lower in the DIG
than in HII regions, indicating that some leakage is necessary to
explain the DIG. However, the difference is not so great that field
stars can be neglected as a source of ionization. The N$_{Lyc}$/\lha~
in the DIG is about 37\% of the ratio in HII regions.  If we normalize
the N$_{Lyc}$/\lha~ to 100\% in HII regions, this relative approach
would suggest that field stars can ionize at least 37\% of the DIG,
well within the uncertainty of our absolute determination.

In the regions covered in the WFPC2 pointings, the fraction of the
{\it total} \ha luminosity that comes from the DIG is 40\%, which is
also the diffuse fraction of the entire galaxy. In the five fields
analyzed here, on average the field stars can account for
40\%$\pm$12\% of the ionization of the DIG (it is a potentially
confusing coincidence that these two numbers are the same).  This
implies that only 30\% of the ionizing photons emitted in HII regions
need to escape to account for the remaining DIG, or put differently,
the predicted \lha~ in HII regions should be 143\% of the observed
\lha~ in HII regions. The average ratio of predicted to observed \lha~
for HII regions is 107\%$\pm$26\%, so the amount of excess ionizing
photons is not enough to explain all of the remaining DIG, with the
maximum within the uncertainty being 133\%. Simply adding up the
observed and predicted \lha, we find that, within the uncertainty,
98\% of the total observed \lha~ (DIG+HII) can be explained by field
stars plus leakage (taking the maximum predicted N$_{Lyc}$ within the
uncertainty). Also keep in mind that other processes may play some
role in ionizing the DIG, such as turbulent mixing layers
\citep{ssb93} and shock ionization. Most likely a combination of these
processes, plus photons leaking from HII regions, ionize the rest of
the DIG. 

There is also the possibility that a fraction of the ionizing photons
are absorbed by dust. \cite{mw97} found that about 25\% of the
ionizing photons emitted by stars within HII regions are absorbed by
dust. If this is the case, we then predict only 80\% of the ionizing
photons necessary to explain the \lha~ of the HII regions. Since there
is not even enough to explain the \lha~ from HII regions, we cannot
then explain the remaining DIG ionization with leaky HII
regions. However, since we know that the HII regions are ionized by
the stars within, the discrepancy might be explained by a systematic
error in determining the spectral types and ionizing luminosities from
the photometry. In this case we could then scale the predicted HII
region luminosity up to 100\%, and scale up the predicted DIG
luminosity accordingly to perform a relative comparison. We could not
address the question of whether any excess ionizing luminosity from
stars in HII regions exists. If an equal fraction of the ionizing
photons emitted by field OB stars is absorbed by dust, the results of
this paper would not be greatly affected.

Taking the maximum predicted N$_{Lyc}$ within the uncertainty, we find
that there are no ionizing photons from stars left over to escape the
galaxy altogether. Of course this depends on the contribution of other
ionization sources to the DIG. If 20\% of the DIG is ionized by
another source, there can be as much as 4\% of the stellar ionizing
photons left over to escape the galaxy, within the error bars. The
uncertainty in these numbers is such that we cannot place much weight
on this limit. However, it is in agreement with \cite{lfhl95}, who found
that less than 3\% of the ionizing photons escape from four starburst
galaxies observed with HUT, although \cite{hjd97} suggest that the
fraction could be as high as 57\% for one of the galaxies by allowing
for absorption by undetected components of the ISM. \cite{d97} also
find very little leakage, less than 1\%. We must also remember that we
are restricted to relatively small regions (about 650pc across), and
are really covering only a small fraction of the disk of M33. We have
know way of knowing whether excess photons emitted by stars in the
image actually escape the galaxy, or simply ionize gas outside of the
image.  Similarly, there may be photons from outside of the image
ionizing gas in the image.

Note that leaky HII regions can provide less than 60\% of the
ionization of the DIG in the regions of M33 studied here. This is in
disagreement with the results of \cite{ok97}, who found that the
Lyman continuum escaping from HII regions could very likely account
for all of the DIG emission in the LMC. This may point to a difference
between the DIG in irregular galaxies and spirals. \cite{mk97} found
that Helium is completely ionized in the DIG in several irregular
galaxies, indicating ionization by very massive stars, type O7 or
hotter. They concluded that density-bounded HII regions are the
dominant source of ionization in those galaxies. A difference in the
ionization source might explain the lower HeI 5876\ang/\ha seen in
some spirals \citep{rt95,r97}.

The field OB stars in M33 appear to have a steeper IMF than OB stars
in HII regions, confirming the results of \cite{m95a} for field OB
stars in the LMC and SMC. The slope which best fits our observations
is not as steep as that found in the LMC and SMC, but it is still
different from that found in HII regions. The difference may imply a
different formation mechanism for field stars, but does not
necessarily mean that they form in the field. Field OB stars may have
formed in an HII region and then drifted out of the dense cloud, or
they may be the remnants of an OB association after the surrounding
gas has dissipated through the actions of SNe and stellar winds. In
either case, lower mass OB stars are more likely to become field
stars, as they have longer lives and thus can drift farther, or live
long enough to outlast the HII region. The result would be a steeper
mass function for field OB stars, because the youngest and most
massive stars are still associated with HII regions. In this case the
varying IMF is used to simulate a steeper {\it mass function} for
stars in the field, and it is not necessary that they have a different
{\it initial} mass function. The steeper IMF merely represents the
fact that the most massive HII region stars have a smaller chance of
becoming field stars due to their short lifetimes. However,
\cite{m95a} carefully corrected for stars which may have drifted out
of HII regions, and found stars as massive as 85 M$_{\odot}$ in the
field. It is difficult to explain how a star this massive became a
field star if it did not form in the field.

\acknowledgments

We thank the anonymous referee for a very careful reading of the
manuscript and comments which improved the presentation of our
results. We are grateful to Bruce Greenawalt for obtaining the \Ha
data, and to Richard Rand for the use of his narrow band filters. The
archival UIT image was obtained through the NASA Data Archive and
Distribution Service. We would like to acknowledge the UIT project for
making their data available. The availability of the Starburst99
models of C. Leitherer and collaborators is greatly appreciated. This
research benefitted from helpful discussions with Jon Holtzman, Salman
Hameed, Bruce Greenawalt, David Thilker, Nichole King, and Vanessa
Galarza. Support for this work was provided by NASA through grant
number AR07645.01-96A from the Space Telescope Science Institute, by
NASA grant NAG5-2426, a Cottrell Scholar Award from Research
Corporation, and by the NSF through grant AST-9617014. CGH was
supported by a grant from the New Mexico Space Grant Consortium.

\newpage 

\begin{figure}
\figurenum{1}
\epsscale{1.0}
\plotone{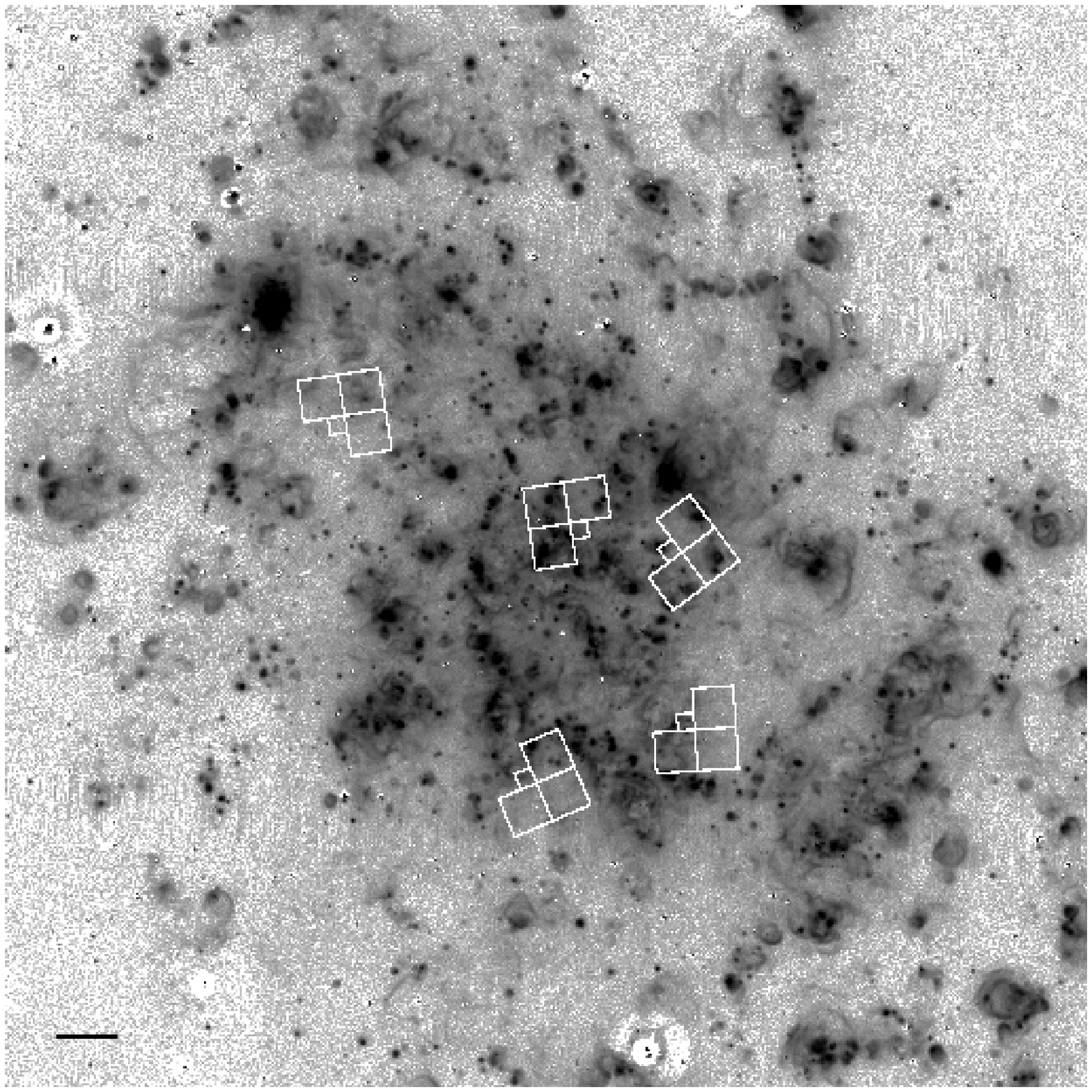}
\caption{Continuum-subtracted \Ha image of M33,
taken with the Burrell Schmidt. North is up and east is to the
left. Only the central 30$^{\prime}$ of M33 are shown in this
figure. The bar in the lower left corner represents 2$^{\prime}$ (=489
pc at the assumed distance of 0.84 Mpc to M33). The overlay shows the
locations of the archival HST WFPC2 data used for this
project.}
\end{figure}

\clearpage
\begin{figure}
\figurenum{2}
\epsscale{0.7}
\plotone{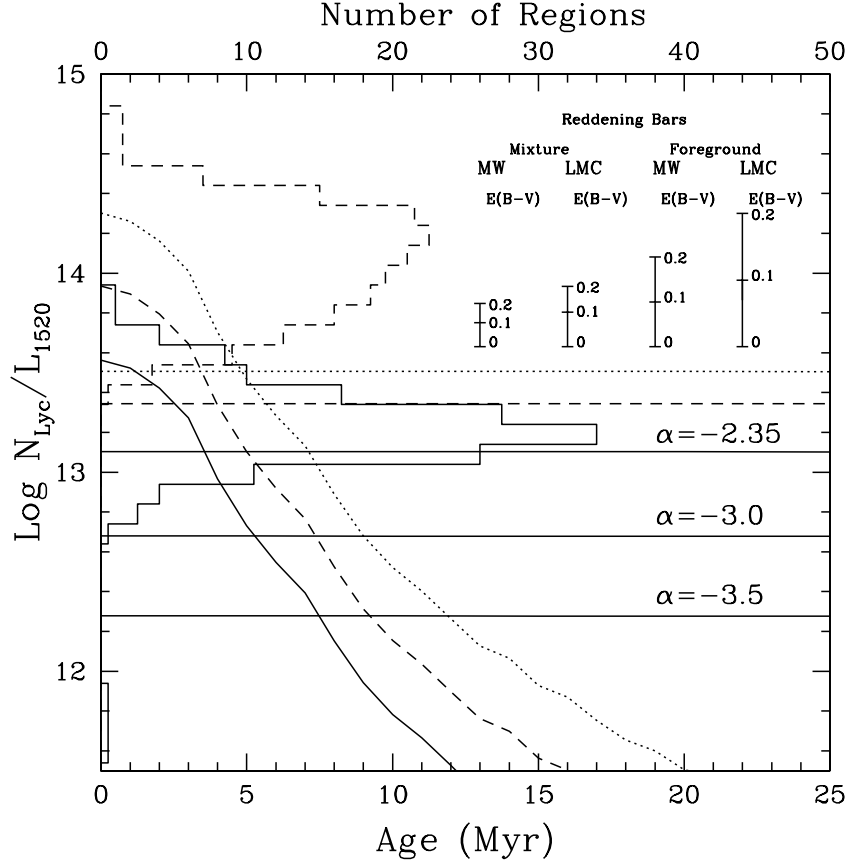}
\caption{Ratio of the number of ionizing photons
(N$_{Lyc}$) to the FUV luminosity from the UIT B1 band image
(L$_{1520}$) for DIG (solid histogram) and HII regions (dashed
histogram). The histograms have not been corrected for reddening. The
models are computed from the Starburst99 evolution code (Leitherer
\etal 1999), with an upper mass cutoff of 100 M$_{\odot}$ and LMC
(Z=0.008) metallicity.  Two models are shown, a single burst (falling
lines) and a steady state model with a constant star formation rate
(straight lines). The dashed lines show the models reddened assuming a
uniform mixture of gas and dust, and the dotted line shows the models
reddened assuming a foreground screen of dust. The burst models were
reddened by E(B-V)=0.2 internal extinction using the LMC extinction
law (Howarth 1983), plus E(B-V)=0.03 Galactic foreground
extinction. The steady state models were reddened by E(B-V)=0.1
internal extinction using the LMC extinction law, plus E(B-V)=0.03
Galactic foreground extinction. The effects of reddening on the models
are shown in the reddening bars at the top right. Predictions for
the steady state model using steeper IMF slopes are also
shown, but without reddened versions.}
\end{figure}

\clearpage
\begin{figure}
\figurenum{3}
\epsscale{0.8}
\plotone{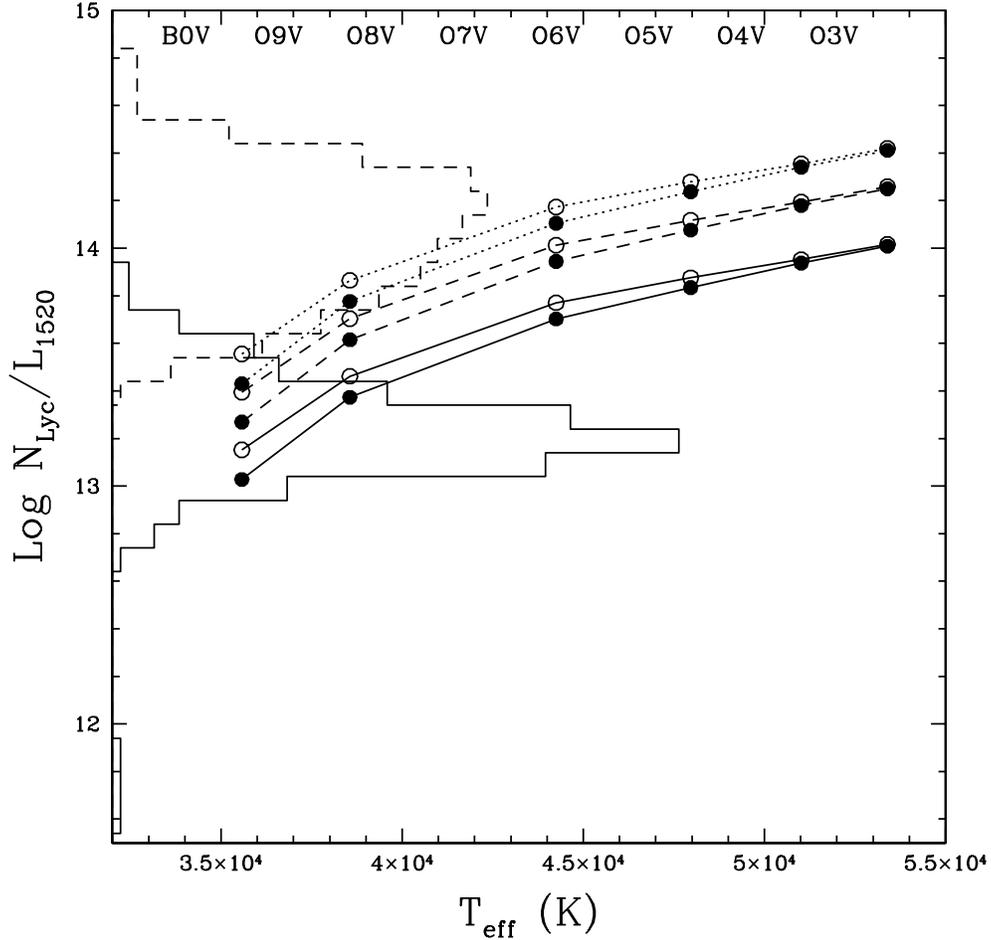}
\caption{Ratio of the number of ionizing photons
(N$_{Lyc}$) to the FUV luminosity from the UIT B1 band image
(L$_{1520}$) for DIG and HII regions (same as figure 2), compared to
model stellar atmospheres. The theoretical ratios were computed using
the CoStar stellar atmosphere models of Schaerer \& de Koter
(1997). The open circles are for models with metallicity Z=0.004, and
the filled circles are models with Z=0.020. The dashed line represents
model stars that have been reddened by E(B-V)=0.1 using the LMC
extinction law and uniform mixture model, and the dotted line
indicates models that have been reddened by E(B-V)=0.1 using the LMC
extinction law and a foreground screen model. Both of the reddened
models also have been reddened by the E(B-V)=0.03 Galactic
foreground.}
\end{figure}

\clearpage
\begin{figure}
\figurenum{4}
\epsscale{0.8}
\plotone{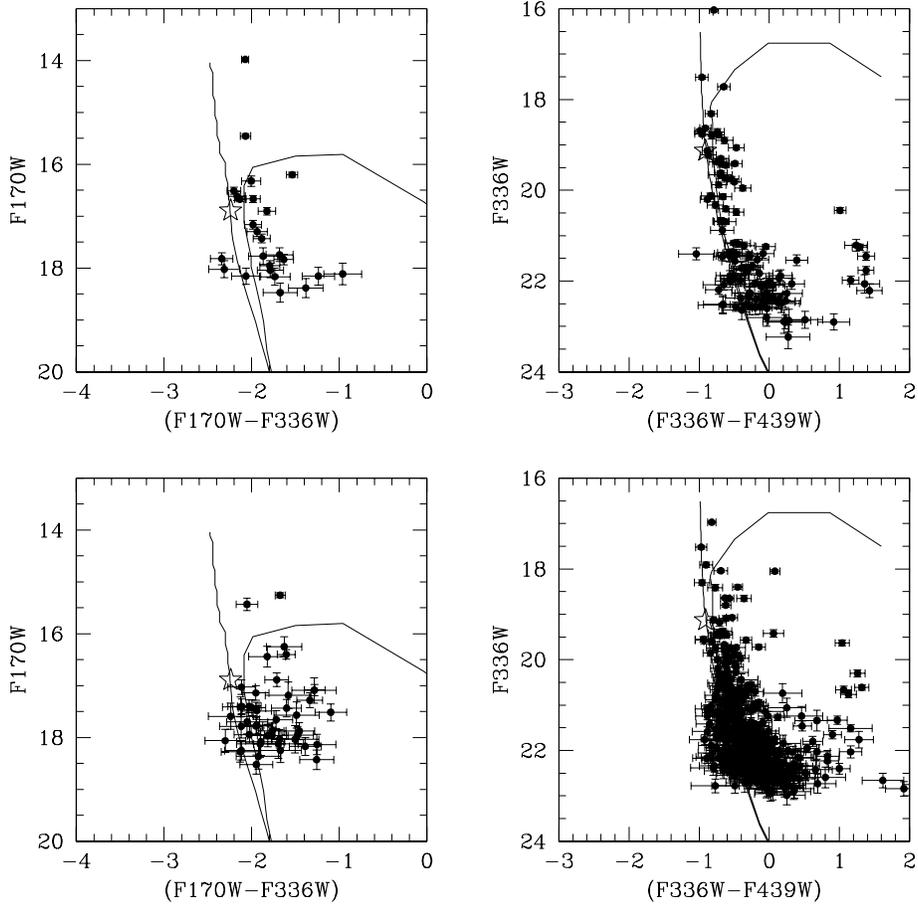}
\caption{Examples of color-magnitude diagrams from the WFPC2
photometry. An HII region is shown in the top two panels, and a region
of DIG is shown in the bottom two panels. Isochrones are from the
Geneva group (Schaerer \etal 1993) and have been converted to the HST
filter system, and show ages of 0 Myr and 7 Myr. They are shown as a
reference point only; they are not the best match to the age of the
population.  The stellar magnitudes have been corrected for reddening
using the LMC extinction law. The amount of reddening was found by
moving the main sequence to fit the isochrones. The 5 pointed symbol
shows the position of a B0V star.}
\end{figure}

\clearpage
\begin{figure}
\figurenum{5}
\epsscale{0.8}
\plotone{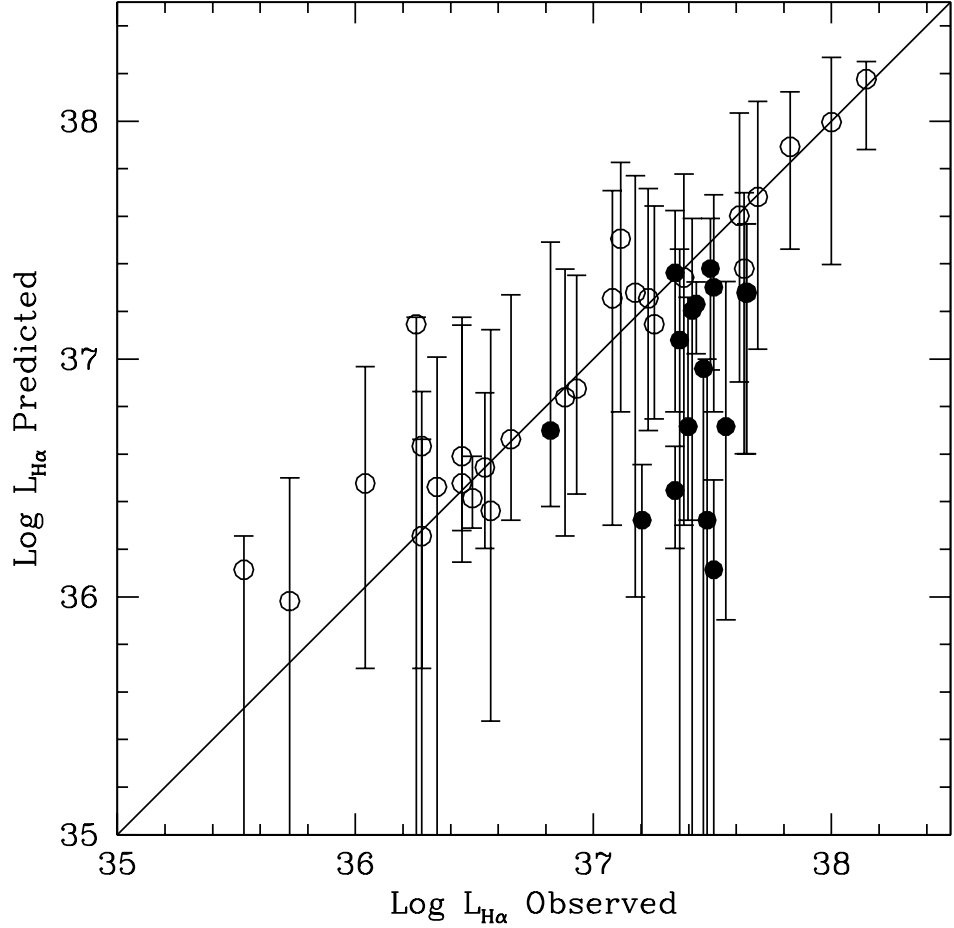}
\caption{Comparison of predicted to observed \ha
luminosity, based on the spectral types of the stars observed in each
region. HII regions are shown as open circles and DIG regions are
shown as filled circles. The error bars are a combination of
uncertainty in the spectral classification due to photometric errors
and uncertainty due to extinction. In some regions the lower errorbar
equals zero ionizing photons.}
\end{figure}

\clearpage
\begin{figure}
\figurenum{6}
\epsscale{1.0}
\plotone{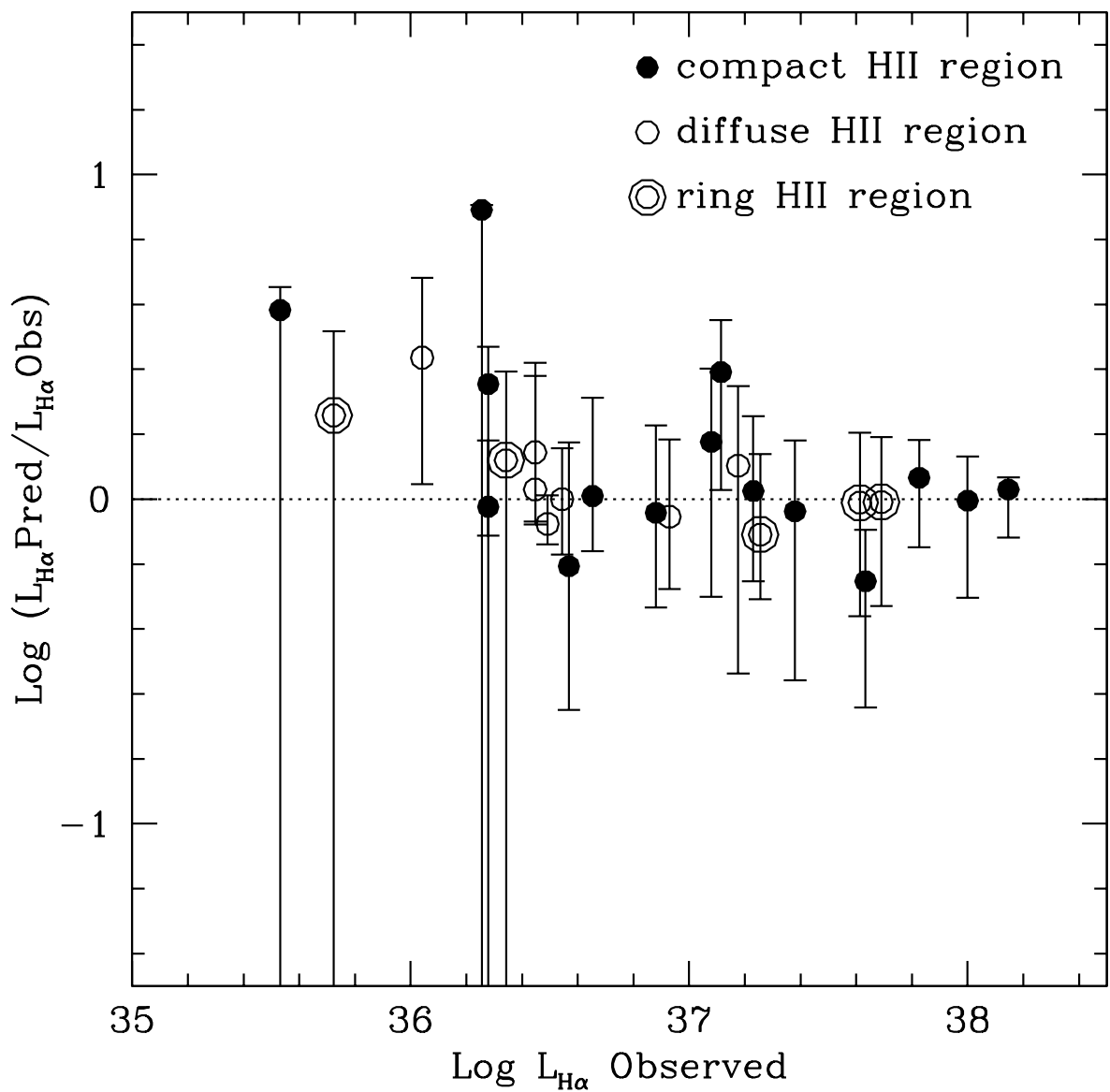}
\caption{Leakiness of HII regions as a function of
luminosity. The points are coded by their morphology, with compact HII
regions shown as filled circles, diffuse HII regions shown as open
circles, and ring-like HII regions shown as double
circles.}
\end{figure}

\begin{figure}
\figurenum{7}
\epsscale{0.8}
\plotone{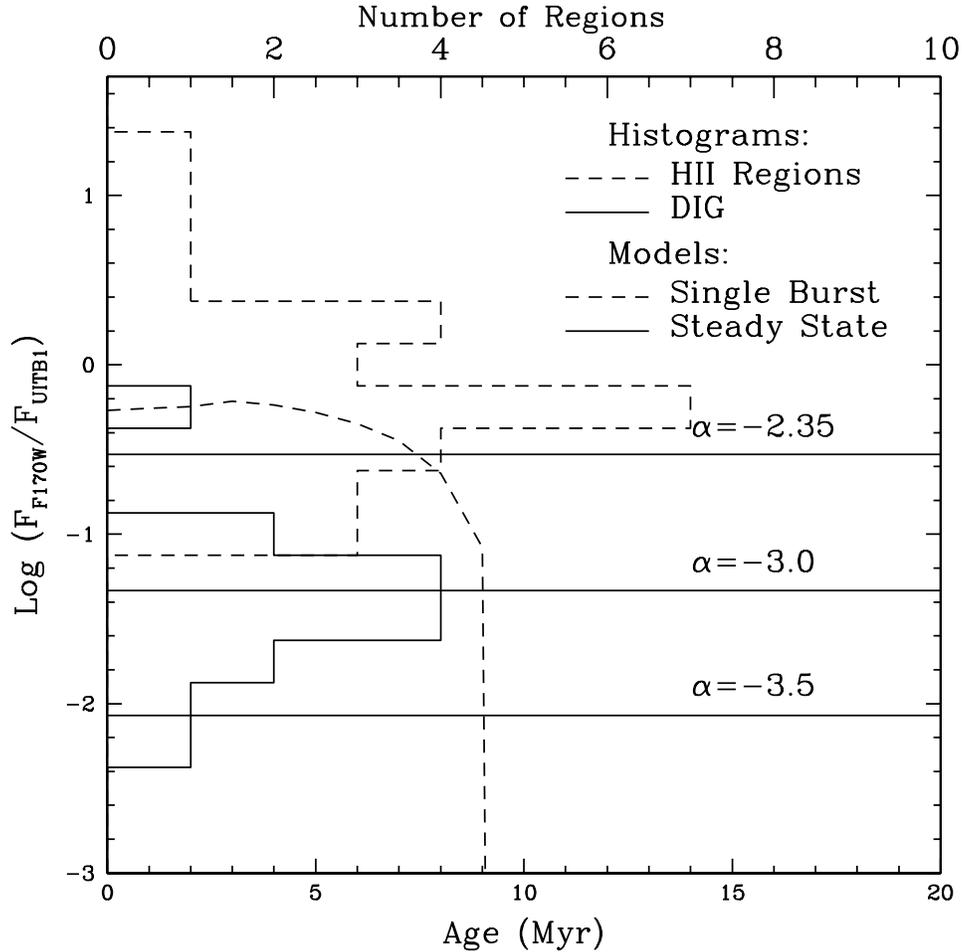}
\caption{Histograms of the FUV flux measured by HST
(F$_{F170W}$)and by UIT (F$_{UITB1}$), compared to models of the
evolution of this ratio with time. The ratio compares the FUV flux
from stars of approximately type B0 and earlier (measued in the HST
images) to the total FUV flux in the region (measured in the UIT
image). HII regions are shown by the dashed histogram, and DIG regions
by the solid histogram. The models are from the Starburst99 models
(Leitherer \etal 1999), and show the behavior of the ratio over time
for a single burst model using a Salpeter IMF with slope
$\alpha$=$-$2.35 (dashed line), and the equilibrium ratio for a steady
state model with a constant star formation rate (solid line), using
IMF slopes $\alpha$=$-$2.35, $-$3.0, and $-$3.5 for massive stars
(these models still use $\alpha$=$-$2.35 for stars with
M$<$20M$_{\odot}$). The HST/UIT ratio in the DIG may reflect a steeper
IMF for massive stars in the field.}
\end{figure}

\clearpage
\begin{deluxetable}{lcccccccc}
\tablecolumns{9}
\tablewidth{0pc}
\tablenum{1}
\tabletypesize{\small}
\tablecaption{Summary of the Data} 
\tablehead{
\colhead{Band} & \colhead{Central $\lambda$} & \colhead{FWHM} & \colhead{Telescope} & \colhead{Exposure Time}\\ 
\colhead{} & \colhead{(\AA)} & \colhead{(\AA)} & \colhead{} & \colhead{(seconds)}} 
\startdata
\Ha             & 6570 & 30   & Burrel Schmidt & 20$\times$900 \\
Off-band        & 6653 & 68   & Burrel Schmidt & 20$\times$540 \\ 
FUV (B1 filter) & 1520 & 354  & UIT            & 424 \\ 
F170W           & 1689 & 434  & HST WFPC2      & 2$\times$900 \\ 
F336W           & 2924 & 727  & HST WFPC2      & 2$\times$900 \\
F439W           & 4292 & 464  & HST WFPC2      & 600 \\ 
F555W           & 5252 & 1222 & HST WFPC2      & 160 \\
\enddata
\end{deluxetable}

\begin{deluxetable}{lcccc}
\tablecolumns{5}
\tablewidth{0pc}
\tablenum{2}
\tabletypesize{\small}
\tablecaption{Comparison of Photometric and Spectroscopically Determined Spectral Types\tablenotemark{a}} 
\tablehead{
\colhead{Spectral Type} & \colhead{Effective Temperature} & \colhead{log $g$} & \colhead{Mass} & \colhead{Log $Q_0$} \\
\colhead{} & \colhead{(K)} & \colhead{(cgs)} & \colhead{(M$_{\odot}$)} & \colhead{s$^{-1}$}}
\startdata
\cutinhead{UIT240 O6-8If\tablenotemark{b}}
Best Match\tablenotemark{c} &   39345  &  3.62 & 60.8  &  49.65 \\
O6Ia       &   41710  &  3.69 & 74.7  &  49.81 \\
O7Ia       &   38720  &  3.58 & 64.3  &  49.69 \\
O8Ia       &   35730  &  3.46 & 54.8  &  49.54 \\
\cutinhead{UIT104 Ofpe/WN9\tablenotemark{b}}
Best Match\tablenotemark{c} &   32257  &  3.27 & 49.8  &  49.31 \\
O8.5Ia     &   34230  &  3.40 & 50.6  &  49.45 \\
O9Ia       &   32740  &  3.33 & 46.7  &  49.33 \\
O9.5Ia     &   31240  &  3.27 & 43.1  &  49.17 \\
\enddata
\tablenotetext{a}{Stellar properties from \citet{vgs96}.}
\tablenotetext{b}{Identification and spectral type from \citet{m96}.}
\tablenotetext{c}{Best match to the CoStar models \citep{sd97}.}
\end{deluxetable}

\begin{deluxetable}{lccc}
\tablecolumns{3}
\tablewidth{0pc}
\tablenum{3}
\tabletypesize{\small}
\tablecaption{Global Ratios} 
\tablehead{\colhead{} & \colhead{L$_{H\alpha}$(Obs)\tablenotemark{a}}& \colhead{L$_{H\alpha}$(Pred)/L$_{H\alpha}$(Obs)}& \colhead{No. of Regions}\\
\colhead{} & \colhead{(erg s$^{-1}$)}& \colhead{}& \colhead{}
}
\startdata
\cutinhead{Four filters required}
All HII Regions& 5.85$\times$10$^{38}$  &1.07$\pm$0.26 & 27 \\
HII L$_{H\alpha}\le$ 5$\times$10$^{36}$ erg s$^{-1}$  &  3.02$\times$10$^{37}$    & 1.60$\pm$0.60 & 13\\
HII 5$\times$10$^{36}<$ L$_{H\alpha}$ $\le$ 5$\times$10$^{37}$ erg s$^{-1}$ & 2.48$\times$10$^{38}$ & 1.01$\pm$0.41 & 11\\
HII L$_{H\alpha}$ $>$ 5$\times$10$^{37}$ erg s$^{-1}$  &   3.07$\times$10$^{38}$    & 1.07$\pm$0.35 & 3\\
All DIG Regions& 4.02$\times$10$^{38}$ &  0.40$\pm$0.12 & 15 \\

\cutinhead{Optical filters only (UBV)}
All HII Regions& 5.83$\times$10$^{38}$ & 1.80$\pm$0.40 & 26 \\
HII L$_{H\alpha}$ $\le$ 5$\times$10$^{36}$ erg s$^{-1}$  & 2.84$\times$10$^{37}$      & 3.66$\pm$2.13 & 12\\
HII 5$\times$10$^{36}<$ L$_{H\alpha}$ $\le$ 5$\times$10$^{37}$ erg s$^{-1}$ & 2.48$\times$10$^{38}$ & 2.39$\pm$0.66 & 11\\
HII L$_{H\alpha}>$ 5$\times$10$^{37}$ erg s$^{-1}$ & 3.07$\times$10$^{38}$   & 1.15$\pm$0.49 & 3\\
All DIG Regions& 4.02$\times$10$^{38}$ & 0.89$\pm$0.22 & 15 \\

\enddata
\tablenotetext{a}{Corrected for extinction.}
\end{deluxetable} 

\end{document}